\documentclass[aps,prl,twocolumn,showpacs,preprintnumbers,amsmath,amssymb]{revtex4-1}
\usepackage{graphicx}
\graphicspath{{Abbildungen/}}
\usepackage{dcolumn}
\usepackage{bm}
\usepackage[mathlines]{lineno}
\usepackage[urlcolor=blue, colorlinks=true]{hyperref}
\usepackage{textcomp}
\usepackage{units}
\usepackage{hyperref}
\usepackage{amsmath}
\usepackage{multirow}
\usepackage{rotating}

\newcommand{\cor}[1]{{\color{black} #1}}  

\usepackage{color}
\newcommand{\COMMENT}[1]{\textcolor{black}{{#1}}}  

\begin{document}
\title{Impact of screening and relaxation onto weakly coupled 2D heterostructures}

\author{T.T. Nhung Nguyen$^{1}$,  T. Sollfrank$^{1}$, and  C. Tegenkamp$^{1}$} \email{christoph.tegenkamp@physik.tu-chemnitz.de}
\affiliation{$^1$ Institut f\"ur Physik,Technische Universit\"at Chemnitz, Reichenhainer Str. 70, 09126 Chemnitz, Germany}

\author{E. Rauls$^2$ and U. Gerstmann$^3$} \email{uwe.gerstmann@upb.de}
\affiliation{$^2$ Department of Mathematics and Physics, University of Stavanger, Norway}
\affiliation{$^3$ Theoretische Materialphysik, Universit\"at Paderborn, D 33098 Paderborn, Germany}
\date{\today}

\begin{abstract}
The stacking of different 2D materials provides a promising approach to realize new states of quantum matter. 
In this combined scanning tunneling microscopy (STM) and density functional theory (DFT) study we show that the structure
in weakly bound, purely van der Waals (vdW) interacting systems is strongly influenced by screening and relaxation.
We studied in detail the physisorption of  lead phthalocyanine (PbPc) molecules on epitaxial monolayer graphene on SiC(0001)
as well as on highly ordered pyrolytic graphite (HOPG), resembling truly 2D and anisotropic, semi-infinite  3D supports.
Our analysis demonstrates that the different deformation ability of the vdW coupled systems, i.e. their actual thickness
\cor{and buckling}, triggers the molecular morphology and exhibits a proximity coupled band structure. 
It thus provides important implications for future 2D design concepts.
\end{abstract}
\maketitle


Heterostructures made layer by layer in a precisely chosen sequence out of 2D materials were suggested  to design bulk quantum materials with entirely new functions \cite{Geim2013}. 
Indeed, proximity coupling reveals superconductivity in twisted bilayer graphene \cite{Novoselov2019, Cao2018}. The absence of dangling bonds in 2D materials is expected to allow a flexible and \cor{lego-like} epitaxial growth of lattice mismatched materials in random order \cite{Koma1999}.

However, as fabricating a 3D stack out of  2D sheets, the same layers may experience a
different coupling, e.g. due to modified screening. Coulomb interaction 
in 2D and 3D is fundamentally different \cite{Cudazzo2011}.
In contrast to the isotropic 3D case,  
for 2D the charge is redistributed on a circle around the point charge, i.e. the residual electric field depends
on the polar angle, resulting in a non-local screening behavior which leads usually to strong and $k$-dependent
renormalization of quasiparticle energies, e.g. excitons \cite{Qiu2013, Ugeda2014, Qiu2017} and
reduced energy gaps \cite{Neaton_2006,Garcia_2009,Noori_2019}.

Among thousands of feasible 2D materials \cite{Mounet2018}, graphene is still the most perfect and flexible one,
thus ideal to elucidate principles of proximity coupling. 
Epitaxial graphene on SiC(0001) provides the flexibility to control the interface and its electronic properties
\cite{Riedl2010, Yazdi2016, Kruskopf2016}: monolayer graphene (MLG) grown on SiC(0001) is $n$-type doped while
quasi-free monolayer graphene (QFMLG) on the same substrate is slightly $p$-type doped \cite{Riedl2009a}. HOPG,
in contrast, is charge neutral and represents the semi-infinite 3D counterpart of graphene \cite{Mammadov2017}.

\cor{Long-ranged ordered molecular 2D structures can be realized also by} physisorption of $\pi$-conjugated
organic molecules on surfaces \cite{Gottfried2015}. \cor{Their combination with solid state 2D
structures proposes advanced stacking sequences with tailored properties.}
However, the comprehensive understanding of the physisorption process can become a formidable challenge.
Usually, the adsorbate layer and surface lattice are not commensurate. Long-range dispersing forces between the
molecules provide a possibility for various phases. The complex interaction scheme with the substrate often
\cor{comes along with charge transfer (between substrate and adsorbate) superimposing the effect of screening}.

Here, the shuttlecock-like lead phthalocyanine (PbPc) molecule with a \cor{large} gap between the highest occupied and
lowest unoccupied molecular orbitals, HOMO and LUMO, 
\cor{helps to suppress} charge
transfer with the substrate. It thus provides an excellent candidate to study implications of screening and proximity
coupling in physisorbed systems. It contains four benzene-pyrrole moieties, which are connected via meso-aza nitrogens. 
The central Pb atom is coordinated to the four adjacent pyrrole nitrogens and is located outside the molecular plane
further reducing the interaction with the substrate.

In this Letter, we analyzed the adsorption of PbPc 
on variously doped epitaxial graphene and HOPG.
For QFMLG and MLG, the \cor{buckling} of the graphene layer \cor{promotes} a quasi-free, densely packed
and chiral PbPc \cor{molecular} layer structure with almost identical lattice parameters.
Only minor 2D screening is observed, so that the molecular states remain almost unaffected.
\cor{In contrast,} large {\em substrate induced} dispersion of the HOMO is found on HOPG, where the more distant molecules interact via
the substrate predominantly. Here, \cor{$\pi$-stacking leads to} proximity coupling of PbPc with \cor{deeper}
graphite layer\cor{s} \cor{and} to a strongly $k$-dependent reduction of the molecular gap.

\vspace*{1mm}

%
\begin{figure}[ttt]
  \centering
        \vspace*{-3mm}
	\includegraphics[width=\linewidth]{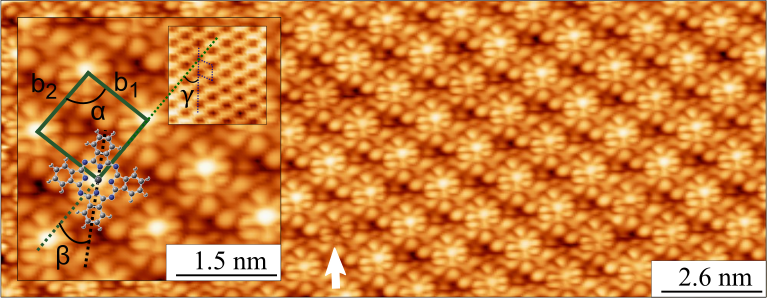}
        \vspace*{-6mm}
	\caption{ Large scale STM image of a \cor{densely-packed molecular} layer of PbPc on
	  QFMLG (0.2~nA, +2~V). The imperfection  (marked by an arrow) is most likely a Pc molecule with a missing Pb atom. The unit cell and relative orientation of the PbPc molecule \COMMENT{w.r.t.  graphene (small inset)} are shown in the inset. The parameters are reported in Tab.~\ref{table}.}
	\label{FIG1}
        \vspace*{-2mm}
\end{figure}

\begin{table}[bbb]
	\vspace*{-3mm}
	\caption{\label{table} Lattice parameters from STM and DFT for \COMMENT{molecular PbPc layers} on HOPG and (QF)MLG as well as
          for a freestanding PbPc \COMMENT{layer}. b$_1$ and b$_2$ denote the lattice constants as shown in Fig.~\ref{FIG1}.
	  $\alpha$, $\beta$   and $\gamma$ denote the angles of the unit cell, the rotation of the molecule and the
          orientation of the unit cell w.r.t. the graphene lattice, respectively. The \cor{minimum} heights of adsorption
          $d$ for atomic type (C/N/Pb) are indicated in Fig.~\ref{FIG3}, \cor{together with the tilting angle $\vartheta$ of
          the molecules and the buckling $\epsilon$ of the topmost C layer.} $N$ denotes the number of C atoms per layer.
          E$_{b/{\rm C}}$, E$_{b/{\rm PbPc}}$ and E$_{b,{\rm intra}}$ refer  to the binding energies per
          substrate C atom, per molecule and the \COMMENT{intra molecular layer}  contribution.}\vspace*{1mm}
	\begin{ruledtabular}
		\begin{tabular}{lc lll}
			& free layer & MLG & QFMLG & HOPG\\	\hline
			b$_1$ (nm) &      &  1.42 $\pm$ 0.05 &  1.40 $\pm$ 0.05 & 1.58 $\pm$ 0.05 \\
			calc.			&  1.35 &   1.37  &  1.37 &  1.54 \\
			b$_2$ (nm)&	  &  1.38 $\pm$ 0.05& 1.40 $\pm$ 0.05 &   1.49 $\pm$ 0.05 \\
			calc.         &  1.33  & 1.30 & 1.30 & 1.54\\
			$\alpha$ ($^{\circ}$) &	  &  90.3 $\pm$ 1.0 & 90.0 $\pm$ 1.0 & 90.1 $\pm$ 1.0 \\
			calc.        & 98.3 & 91.9 & 91.9 & 92.2 \\
			$\beta$ ($^{\circ}$) & --- & 30 $\pm$ 3& 30 $\pm$ 3&30 $\pm$ 3 \\
			$\gamma$   ($^{\circ}$) &    & 39.2 $\pm$ 2.0& 38.5 $\pm$ 2.0 & 30.5 $\pm$ 2.0 \\
			calc. & --- & 39.4 &  39.1  & 30.0 \\[2mm]
			height $d$ (\AA) & --- & 2.5/3.5/4.6 & 2.3/3.5/4.7  &  3.1/3.3/4.7 \\
                        buckling $\epsilon$ (\AA)  & --- &  0.51  &  0.67  &  0.02  \\
                        tilting $\vartheta$ ($^{\circ}$) & --- &  9.2   &  9.5  &   $ 0.2$ \\  
                        $ N_{\rm C/layer}$      & ---  &   68 &   68 &  90 \\
			E$_{b/\rm{C}}$ (meV)     & ---  &   29 &   31 &  26 \\
			E$_{b/\rm{PbPc}}$ (eV)   & 1.18  & 1.97 & 2.11 & 2.35 \\
				E$_{b,\rm{intra}}$ (eV) & 1.18 & 0.25 & 0.24 &  0.01 \\
		\end{tabular}
	\end{ruledtabular}
\end{table}

\begin{figure}[ttt]
  \centering
        \vspace*{-5mm}
	\includegraphics[width=1.0\linewidth]{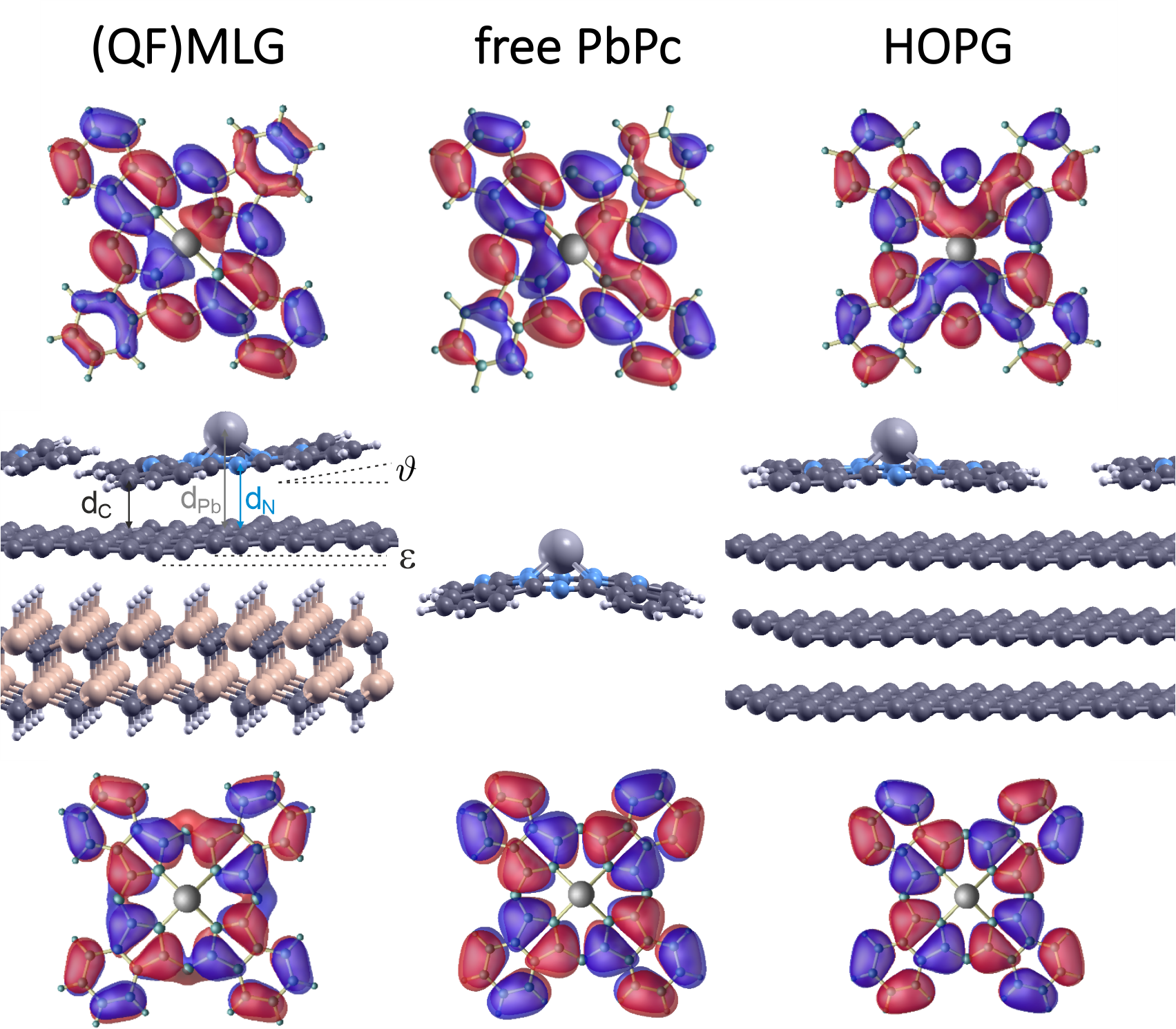}
        \vspace*{-5mm}
        \caption{Structure of PbPc monolayers on (QF)MLG \cor{(tilting angle $\vartheta$)} and HOPG, in comparison with isolated (free) PbPc molecules.
          The molecular HOMOs (bottom) and LUMOs
          (top) for the molecular species alone are also shown. For HOPG, the strong coupling to the other layers prevents the
          topmost layer from corrugation and \cor{buckling $\epsilon$, cf.~Table I.}} 
	\label{FIG3}
\end{figure}

As substrates epitaxial monolayer graphene (MLG) and  hydrogen-intercalated quasi-free monolayer graphene (QFMLG) on
semi-insulating 6$H$-SiC(0001) as well as  highly ordered pyrolytic graphite (HOPG)  were used.                         
While HOPG resembles a charge neutral anisotropic 3D material, MLG and QFMLG are both 2D materials but with
\cor{completely} different electrochemical potentials \cite{Mammadov2017}. 
Details about fabrication \COMMENT{and characterization} of the substrates are reported  elsewhere  \cite{Rield2008,Riedl2009, Momeni2018,Momeni2019}.
The adsorption of PbPc  molecules 
was done at 300~K under ultra high vacuum with same adsorption rates in order to allow a direct comparison \cite{NhungNguyen2019}. 
The atomic structure and positions of molecular energy levels were investigated by low temperature scanning tunneling microscopy
(LT-STM, 6~K and 80~K). Scanning tunneling spectroscopy (STS) was recorded using  lock-in technique (20~meV, 1~kHz). \COMMENT{For the dI/dV spectra an average of at least 10 curves were acquired at various positions across the molecules.}

Our experiments are supplemented by DFT calculations \cor{using a supercell approach and periodic boundary conditions}.
For HOPG, \COMMENT{molecular layers} of PbPc molecules were modeled on 6 layer thick Bernal-stacked graphite.
Thereby, the simplicity of the substrate allows a direct modeling in the experimentally observed quasi-square \cor{surface}
unit cell (with $N_{\rm }$=90 atoms per C layer, see Tab.~\ref{table}).
In contrast, for QFMLG and MLG, square unit cells are either incommensurable with the underlying  SiC(0001) substrate
or the graphene layers. As a result, the unit cell of PbPc  on (QF)MLG/SiC(0001) contains at least two molecules.
However, the absence of any indications in the STM experiment suggests that the SiC part of the substrate plays a minor role.
Thus, the MLG calculations were restricted to a simplified unit cell containing one PbPc, where the interaction with the
substrate (68 atoms per C layer) is reduced to the topmost graphene layer plus a partially H decorated buffer layer,
\cor{whereby the level of doping increases almost linearly with the number of in this way $sp^3$-coordinated C atoms.
  The doping level, \cor{i.e. the position of the Dirac point E$_D$ (cf. Fig. 5a),} can thus be adjusted via the degree
  of H decoration. For a 0.4 eV shift determined experimentally \cite{Momeni2019}, about 15 \% of the C atoms of the buffer
  layer have to be covalently bound to the underlaying SiC substrate (cf. Fig. 5b), in fair agreement with experiment \cite{Emtsev2008}.}
 
Structural relaxation calculations are performed with the Quantum ESPRESSO package using
periodic boundary conditions and a 3$\times$3$\times$1 $k$-point sampling \cite{QE-2009,QE-2017}. STM images are simulated based on
VASP calculations using the Tersoff-Hamann approach to analyse the tunneling current~\cite{Tersoff-Hamann_1985}.  
Specifically, we use scalar relativistic norm-conserving pseudopotentials and a plane wave basis set with 90 Ry energy cutoff.
For structure relaxation the semi-local PBE functional was used to include many-body effects due to exchange and correlation (XC).
Afterwards the B3LYP hybrid functional was used to accurately determine the electronic structure for the PBE relaxed structures.
\cor{The use of B3LYP  copes the DFT-underestimization of the molecular HOMO-LUMO gap (see Ref.~\onlinecite{Baran2010b,Zhang2007}
and  Table II), and allows a 1:1 comparison of the resulting density of states (DOS) with the experimental STS spectra.}
In all calculations the D3 dispersion correction was used for a reasonable description of non-local
correlation effects \cite{Grimme_D3}. 
 

Our experiments and calculations clearly reveal an adsorption of PbPc, where the central Pb atom is pointing upwards on HOPG
as well as on epitaxial graphene, as shown in Fig.~\ref{FIG1}.
At least 0.3~eV per molecule (0.72 eV on HOPG) is gained by this preferential adsorption geometry. 
This is in contrast to Au(111) surfaces, where PbPc molecules show both, up- and down configurations \cite{Madhuri2017}.
PbPc on Cu(100) and Ag(100) have been found to form a chiral monolayer structure, while on Pd(100) a stable {\em achiral}
state was reported \cite{Mugarza2010,Cai2015}.  The latter adsorption geometry was explained by stronger hybridization
between the $p_z$ orbital of the macrocyclic C atoms in PbPc and the $4d$ orbitals of the Pd substrate.

\begin{figure}[ttt]
	\centering
	\includegraphics[width=0.95\linewidth]{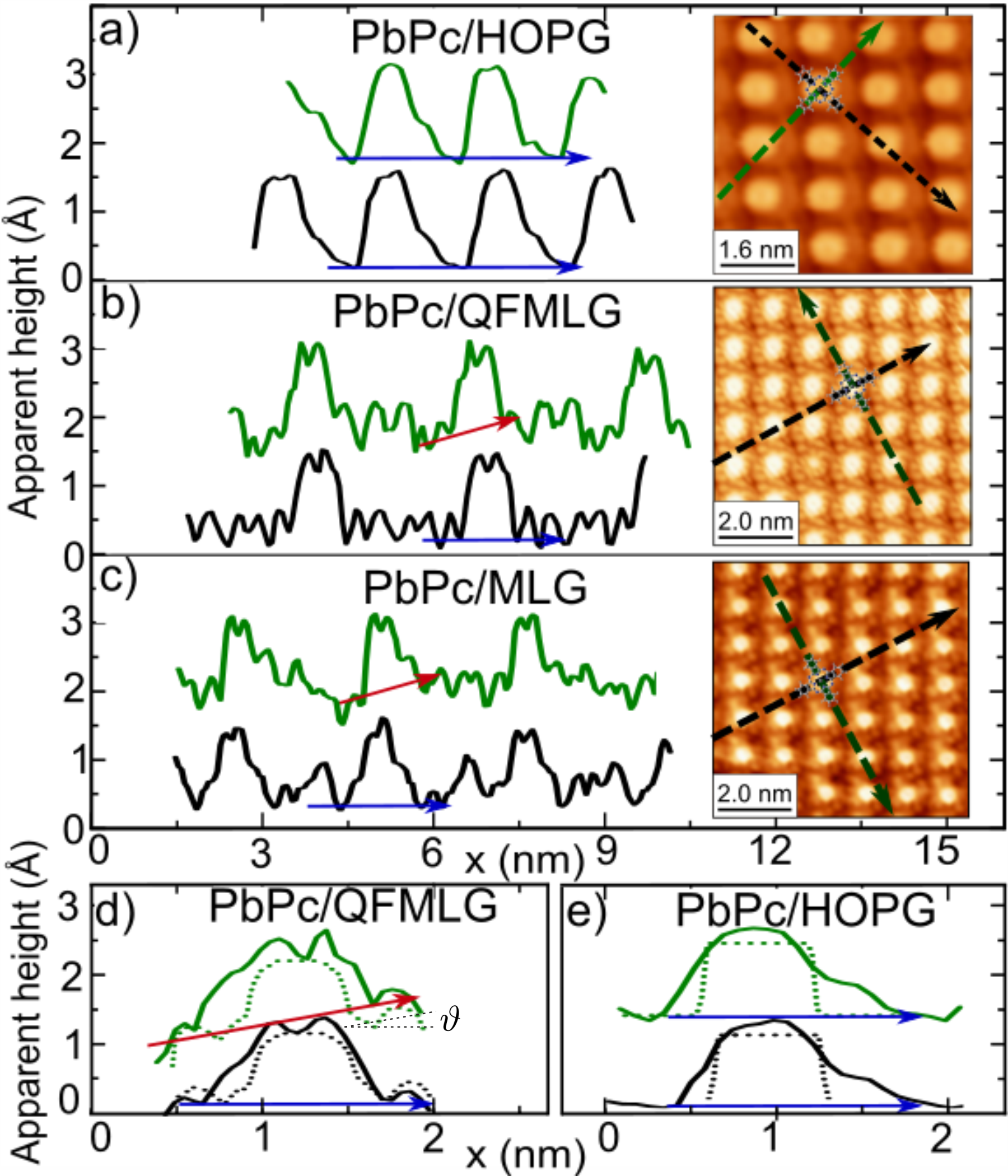}
        \vspace*{-2mm}
	\caption{High resolution STM images of PbPc monolayers on (a) HOPG (0.2~nA, -2~V), (b) QFMLG (0.2~nA, -2~V) and  (c) MLG (0.2~nA, -1~V)  and corresponding height profiles taken along two main molecular axes as indicated. A \cor{tilting} (red arrows) is seen only on (QF)MLG along one direction (green).  d) and e) show magnifications of PbPc/QFMLG and PbPc/HOPG, respectively, together with profiles (dotted lines) obtained from DFT calculated STM images, Fig~\ref{FIG4}.} 
        \vspace*{-2mm}
	\label{FIG2}
\end{figure}

\begin{figure}[ttt]
	\centering
	\includegraphics[width=\linewidth]{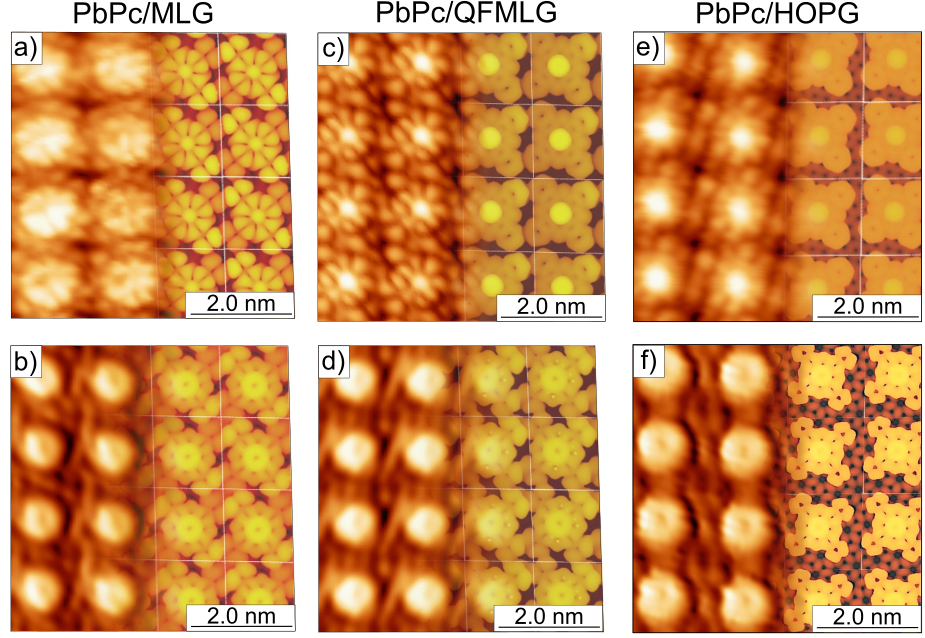}
	\vspace*{-5mm}
	\caption{Comparison of measured (left) and DFT simulated (right) STM patterns for  negative (HOMO, bottom) and positive voltages (LUMO, top), i.e. occupied and empty states. The tunneling conditions, given in (nA/V), were: a) 0.2/+2, b) 0.2/-2, c) 0.2/+2, d) 0.5/-2, e) 0.1/+1.6, f) 0.2/-2, whereby the current was chosen in order to optimize the contrast.}
	\label{FIG4}
\end{figure}

\begin{figure}[ttt]
  \centering
	\includegraphics[width=1.0\linewidth]{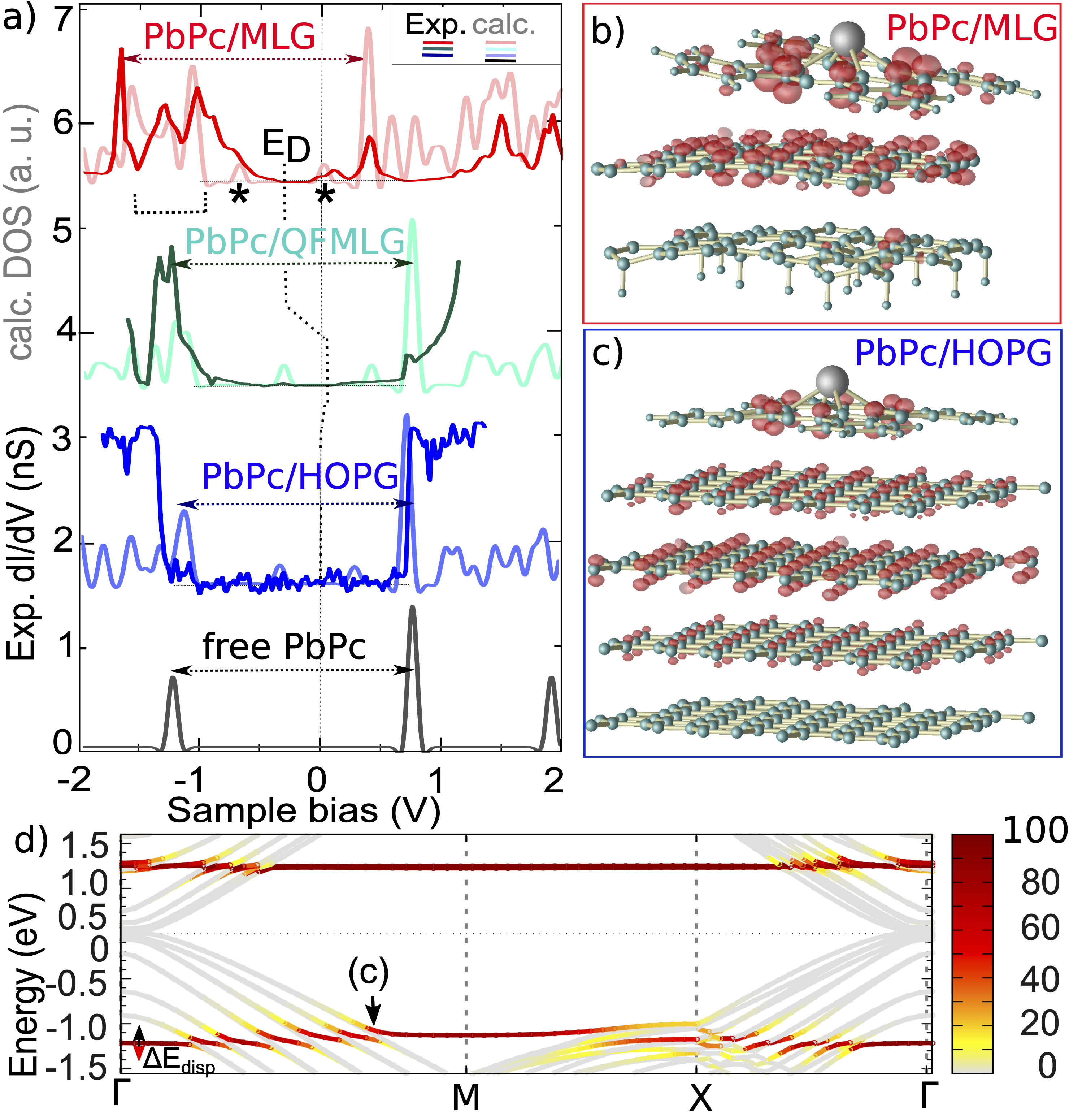}
        \vspace*{-5mm}
	\caption{a) Averaged dI/dV spectra (set-point: 0.2~nA, -1~V) taken for PbPc on MLG, QFMLG and HOPG
          compared with calculated DOS (B3LYP-D3, shifted $3\times3\times3$ $k$-sampling).
          Structure and molecular HOMO interacting with the substrate for MLG (b) and HOPG (c),
          \cor{whereby the spatial distribution of the HOMO (red) is exemplarily shown for a $k$-point
            indicated by the arrow in d): For HOPG,} the $k$-dependent contributions
          of the molecule (in \%) \cor{to the HOMO and LUMO are} indicated by the color scale (d). \COMMENT{The $\pi$-states of the substrates are labeled by ($\ast$).
          Due to finite $k$-sampling, they appear at finite energies (centered around the Dirac point E$_D$) and allow the accurate determination of E$_D$. }} 
	\label{FIG5}
\end{figure}

PbPc on all the investigated graphene and graphite templates forms highly-ordered {\em chiral} monolayer structures with a
single PbPc molecules in quasi-square unit cells, as shown in Fig.~\ref{FIG1} exemplarily for QFMLG~\cite{NhungNguyen2019}.
Table~\ref{table} summarizes the lattice parameters and molecular orientations  
which were deduced from STM images taken across the edges of the PbPc islands (cf. with Ref.~\cite{NhungNguyen2019}).
\cor{They are nicely confirmed and rationalized by our DFT simulations, i.e. by minimizing total energy
while varying cell size and shape.}\\[-3mm]

The lattice parameters for PbPc on QFMLG and MLG are similar (even identical in theory), while the parameters found on HOPG
are considerably larger by about 10\% (cf. Tab.~\ref{table}).
The different lattice parameters come along with specific details of the adsorption structure.
The characteristic shuttlecock structure of the free PbPc relaxes upon physisorption on all the three substrates.
In case of HOPG, the C$_{4v}$ symmetry of the gas phase PbPc molecules is retained, but all wings are found almost
planar, cf.~Fig.~\ref{FIG3}.
It maximizes the attractive vdW interaction {\em per molecule} with the graphene template and resembles the geometry of
isolated physisorbed molecules \cite{Kera2007,Shibuta2010,Yamamoto2011}.
The adsorption height of the C-atoms of about 3.1 \AA\ (cf. Tab.~\ref{table}) is similar to the interlayer distance in
graphite. Together with the planar adsorption geometry this suggests
$\pi-\pi^*$ stacking as a predominant driving force.  
In essence, this stacking 
\cor{gives} rise to a proximity-coupled band structure, as we will show below. 

In contrast, the C$_{4v}$ symmetry of the PbPc upon adsorption on (QF)MLG is lifted. Two neighboring benzene-pyrrole
units are bended towards the surface, while the others are lifted to different extent (cf.~Fig.~\ref{FIG3}).
The formation of a layer of in this way tilted molecules is in agreement with the asymmetry seen in the STM height
profiles taken along the two principal axes of the molecules on (QF)MLG (cf. Fig.~\ref{FIG2} b,c). Details of the
height profiles are shown in the close-up in panel d) and coincide \cor{in all cases} with profiles obtained from
DFT calculated STM images shown in Fig.~\ref{FIG4}.

The tilted structure on (QF)MLG allows a closer arrangement with \cor{strongly} increased \cor{($\times$ 20)}
intermolecular coupling, E$_{b,\textrm{intra}}$ (see Tab.~\ref{table}) while providing a maximum binding energy
{\em per substrate area} (i.e. per C-atom).
\cor{The resulting lattice constants are about 10\% smaller than on HOPG and, notably, comparable
to a potential freestanding molecular PbPc layer}
\footnote{The overall total energy minimum of a free molecular layer is found for a lattice equivalent to a coverage of 1
molecule per 71 C atoms, very close to the (QF)MLG structures (1 PbPc molecule per 68 C atoms, see Table~\ref{table}).}.
\vspace*{1mm}

{\em What is the driving force behind the different adsorption schemes?}
---
The geometry of  PbPc on MLG and QFMLG is very similar, \cor{despite their} different electrochemical potentials.
\cor{Obviously, the doping level} of the two 2D substrates are of minor relevance \cite{Mammadov2017}.
\cor{According to recent transport measurements \cite{NhungNguyen2019}, charge transfer is also not taking place,}
in agreement with \cor{the present} STS and DFT calculations (see below).

A conceivable reason is the corrugation of epitaxial graphene on SiC.
\cor{The buckling} facilitates adsorption of {\em tilted} PbPc molecules:
The topmost graphene layer is considerably upwards bended towards the PbPc macrocycle compensating the tilting-induced
losses in vdW interaction.
Although the exfoliation energy for HOPG is one order of magnitude lower than in case of MLG/SiC~\cite{Wang2015, Wells2014}
a local deformation of the uppermost C layer in HOPG costs by far more energy (182 meV instead of 86 meV for MLG).
Obviously, the inherent corrugation of epitaxial graphene layers (the lateral strain \cor{also responsible for the
buckling}) allows for a flexible adaption of the substrate to the adsorbed molecular structures.
Thus, the tilting of the PbPc molecule is a direct consequence of the 
deformation ability of the 2D support.

A  common STM feature for all three substrates is a donut-like shape of the occupied state in the center of the molecule.
It was seen for all investigated substrates and is nicely reproduced by DFT in Fig.~\ref{FIG4} \cite{NhungNguyen2019}.
\cor{This demonstrates} that the central Pb atom does not contribute to the HOMO (cf. Fig.~\ref{FIG3},~\ref{FIG5}b,c)
\cor{for all ivestigated substrates, whereby} this spectroscopic fingerprint becomes most obvious at slightly \cor{different}
tunneling voltages, see Fig.~\ref{FIG4}. For a more detailed analysis, additional STS measurements were performed. 
In Fig.~\ref{FIG5}~a) averaged dI/dV-spectra are shown and compared with the B3LYP-D3 calculated density of states (DOS):

%
\begin{table}[bbb]
  \vspace*{-5mm}
  \caption{\label{table_2} Calculated  HOMO-LUMO gap energies (eV, B3LYP-D3 hybrid functional) for PbPc on various substrates. For comparison the values for isolated (single) molecules \cor{(also for (semi-)local PBE/LDA functionals)} and freestanding  molecular monolayers (B3LYP-D3) are also \cor{given}. For the periodic structures the $k$-point dependence (12$\times$12$\times$1 sampling) is documented by the minimum/maximum values  and the dispersion
    $\Delta$E$_{\rm disp}$= gap$_{\rm max}$ -- gap$_{\rm min}$.} \vspace*{1mm}
 	\begin{ruledtabular}
	  \begin{tabular}{l c cc}
            System (XC-funct.)&  isolated    &  gap$_{\rm max/min}$ & $\Delta$E$_{\rm disp}$ \\
         \hline\vspace*{-2mm}\\ 
         PbPc isolated (LDA,PZ-D3)&     1.32     &               &         \\
         PbPc isolated (PBE-D3)   &     1.33     &               &         \\
         PbPc isolated (B3LYP-D3) &     2.01     &               &         \vspace*{1.5mm}\\
         freestanding PbPc film (MLG)  \hspace*{-5mm}  &           &   2.03/2.00   &  0.03  \\     
         freestanding PbPc film (QFMLG)\hspace*{-3mm}  &           &   2.01/2.00   &  0.01  \\ \vspace*{1.5mm}
         freestanding PbPc film (HOPG) \hspace*{-4mm}  &           &   2.01/2.01   &  0.00  \\ 
         PbPc/MLG              &     1.98     &   1.98/1.94   &  0.05  \\
         PbPc/QFMLG            &     1.95     &   1.94/1.85   &  0.09  \\
         PbPc/HOPG             &     2.00     &   2.00/1.75   &  0.39  \\
		\end{tabular}
	\end{ruledtabular}
\end{table}

(i) While the energies for the HOMO and LUMO states of gas phase PbPc~\cite{Baran2010b} and adsorbed
on QFMLG \cor{as well as} HOPG are similar, the spectrum
measured on MLG is shifted to lower energies (by $\approx$0.4~eV), reflecting the $n$-type doping of MLG.
\cor{Whereas the position of the LUMO is obvious also in this case, the identification of the HOMO requires
theoretical support:} Those C atoms \cor{(of the buffer layer)} covalently bound to the SiC substrate,
introduce \cor{additional occupied states partially superimposing the HOMO}  (\COMMENT{see bracket in } Fig.~\ref{FIG5}).


(ii) Interestingly, the different doping levels of MLG and QFMLG \cor{play no role}.
Very similar HOMO and only slightly different LUMO STM
signatures \cor{suggest that} both structures experience an almost identical \cor{lateral} screening behavior.
There are also only minor 
relative shifts of the HOMO and LUMO levels
(cf. Tab.~\ref{table_2}). This is in line with literature where relevant \cor{screening}
effects onto the \cor{molecular} electronic structure
are restricted to substrate 2D-layer distances clearly below 3 \AA~\cite{Niesner_2014,MARKS2014263}. 

Contrary, for HOPG our B3LYP-D3 band structure calculations reveal a \cor{large dispersion} of the molecular HOMO ($\Delta E_{disp}$ $\approx$ 0.39~eV).
The LUMO is affected by much lower extend (cf. Fig.~\ref{FIG5}~d).
Similar to the case of metallic substrates \cite{Cai2015,Kera2007}, the
resulting $k$-point dependent \cor{renormalization} of the molecular HOMO-LUMO gap
can be attributed to $\pi$-stacking of the eight macrocyclic C atoms (those bridging two N atoms) with the substrate
C atoms (see Fig.~\ref{FIG5}~b,c) \COMMENT{which is largest for the second graphite layer. Notably, this hybridization effect
is not coming along with a gap opening in the substrate bands and, thus, mimics an example of proximity coupling}. 
The concomitant modification of the band structure is strongly $k$-dependent,
and the maximum size effect is restricted to rather small regions within the Brillouin zone (Fig.~\ref{FIG5}~d),
explaining the shoulder observed in STS \cor{slightly below --1~V sample bias}. 
The fact, that the molecular states of PbPc on HOPG reveal a strongly enhanced dispersion,
although the intermolecular distance is larger compared to (QF)MLG), underlines the importance of a substrate mediated
interaction.

In summary, we comprehensively studied vdW interacting heterostructures by means of PbPc monolayer structures
on 2D graphene, \cor{(QF)MLG,} and semi-infinite 3D graphite, \cor{HOPG}. Albeit the surface structure of all
templates are the same and charge transfer is not taking place, the molecular layer reveals very different
lattice parameters and underwent different relaxation schemes.
Formation of almost identical densely packed PbPc \cor{molecular} layers with strongly tilted molecules were
found on both 2D templates, despite their very different work functions, showing that lateral Thomas Fermi
screening plays a minor role.
Contrary, the interaction with the upper graphite (HOPG) layers, {\em in particular the second}, favors an
almost planar adsorption of PbPc at the expense of a considerably larger lattice constant. The dispersing
molecular states unambiguously demonstrate the presence of a substrate mediated interaction and the band
structure exhibits spectral features of proximity coupling.  
Therefore, the actual thickness of a 3D stack built from 2D sheets appears to be decisive for the
vdW heteroepitaxy and impacts recent layer by layer design concepts \cite{Koma1999,Geim2013}.

\vspace{1ex}
\textbf{Acknowledgement}\\
We thank D. Momeni Pakdehi (PTB Braunschweig) for providing us epitaxially grown graphene samples on SiC(0001).
Financial support by the DFG (through project Te386/17-1 and TRR~142, project number 231447078) is gratefully acknowledged.
The in parts demanding numerical calculations were possible thanks to CPU-time grants of the Paderborn Center for Parallel Computing, (PC)$^2$.
%

\end{document}